# A CONCEPT FOR CANCELLING THE LEAKAGE FIELD INSIDE THE STORED BEAM CHAMBER OF A SEPTUM MAGNET*

M. Abliz†, M. Jaski, A. Xiao, A. Jain, U. Wienands, H. Cease, M. Borland, G. Decker, J. Kerby
Argonne National Laboratory, Argonne, IL 60439, U.S.A

*Abstract*

The Advanced Photon Source is in the process of upgrading its storage ring from a double-bend to a multi-bend lattice as part of the APS Upgrade Project (APS-U). A swap-out injection scheme is planned for the APS-U to keep a constant beam current and to enable a small dynamic aperture. A novel concept that cancels out the effect of leakage field inside the stored beam chamber was introduced in the design of the septum magnet. As a result, the horizontal deflecting angle of the stored beam was reduced to below 1 μrad with a 2 mm septum thickness and 1.06T normal injection field. The concept helped to minimize the integrated skew quadrupole field and normal sextupole fields inside stored beam chamber as well.

The designed septum magnet deflects the injected electron beam by 89 mrad with a ring energy of 6 GeV. The stored beam chamber has an 8 mm x 6 mm super-ellipsoidal aperture. The magnet is straight; however, it is tilted in yaw, roll, and pitch from the stored beam chamber to meet the on axis swap out injection requirements for the APS-U lattice.

## INTRODUCTION

In order to increase the brilliance of synchrotron X-ray, APS is planning to reduce the beam emittance to 41 pm by introducing seven bend achromat concept to its lattice as an upgrade (APS-U) project [1-2]. On axis swap-out injection [3-4] is required for the APS-U. The electron beam trajectory needs to be deflected in the horizontal and vertical planes before it is injected into the storage ring [5]. The septum needs to be tilted in yaw, pitch, and roll in order to make the required on-axis injection, since the beam comes at an angle from the booster.

The stored beam deflection at the exit of the septum magnet should be avoided by controlling the leakage field inside the stored beam chamber. With the original Lambertson septum magnet, the leakage field becomes harder to control if the septum is thin and injection field is high. Moreover, the skew quadrupole field inside the stored beam chamber becomes larger if the leakage field is high. Therefore, the injection field strength was controlled at 0.75 T with a septum thickness of 2.4 mm [6] to avoid those peculiar problems of the Lambertson septum magnet at the current APS.

___________________
* Work supported by the U. S. Department of Energy, Office of Science, under Contract No.
DE-AC02-06CH11357
† email address: mabliz@aps.anl.gov

The APS-U specification for the septum magnet is listed in Table 1. The required septum thickness is 2 mm. The available space limitation of 1.78 m results in a peak field for the injected beam of more than 1 T in order to achieve the total deflecting angle of 89 mrad.

A thin septum with a high injection field makes the design challenging in terms of the deflecting angle (or field leakage) and excessive skew quadrupole field seen by the stored beam [7]. Furthermore, the required super-ellipsoidal cross-section of the stored beam chamber increases the field leakage in the stored beam chamber compared to the commonly used round beam chamber. To address these design challenges, a novel concept that cancels out the leakage field was developed and applied to the APS-U septum design. The concept was placing the US (upstream) end of the stored beam chamber under the side leg to create a positive $B_y$ field leakage to cancel out the negative $B_y$ leakage field that was created at the DS (downstream) end septum thickness. Almost all the integrated leakage field inside stored beam chamber was cancelled out. Furthermore, the concept helped to reduce the integrated skew quadrupole field and normal sextupole also inside the stored beam chamber.

Table 1: Specification

| Parameter | Value | Unit |
|---|---|---|
| Magnet Type | DC | --- |
| Injected Beam Deflecting Angle | 89 | mrad |
| Injected Field Strength, By | 1 | T |
| Roll Angle | 93 | mrad |
| Stored Beam Deflecting Angle | < 100 | μrad |
| Field Uniformity (ΔB/B), +/- 2 mm beam vicinity | ≤ 0.001 | --- |
| Insertion Length | 1.78 | m |
| Aperture Stored Beam Chamber | 8 x 6 | mm |
| Septum Thickness at DS | 2 | mm |
| Septum Thickness at US | 4.56 | mm |
| Beam separation at DS septum | 5.5 | mm |
| Septum Thickness Tolerance | 50 | μm |

The absolute value of the peak leakage field at the septum thickness was reduced dramatically by applying three more useful ideas to the design of the septum magnet: 1) the top pole was cut shorter than the bottom pole at both US and DS ends; 2) an open space was created around the stored beam chamber; 3) Vanadium Permendur (VP) was selected as the material of the stored beam chamber. The detailed design, injection and stored beam trajectories, the field along the injection beam trajectory, and leakage field inside the stored beam chamber with the concept are reported.

## MAGNETIC DESIGN

H-shaped dipole magnet structure was designed for the septum magnet with Opera 3D, as shown in Fig. 1. A 14 - layer coil with 4 turns per layer coil, 4 x 14 turns, was wound around the top pole. The gap between the top and bottom poles was set at 10 mm.

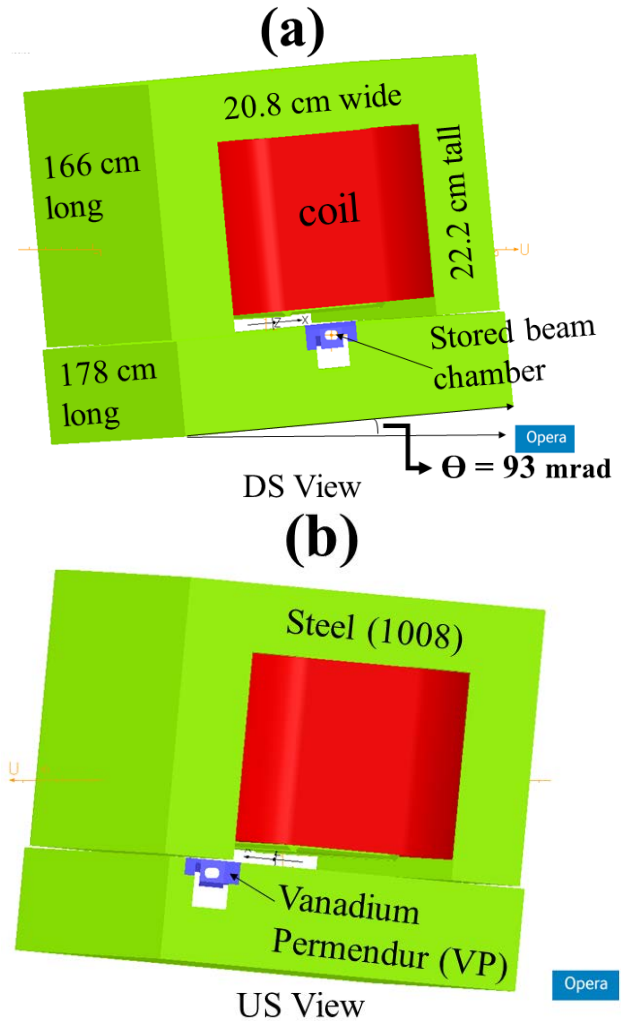

Figure 1: (a) & (b) are the views of the septum magnet from the stored beam chamber at DS and US ends.

The stored beam chamber is located in the bottom pole. The width of the top pole is 6.5 cm, the bottom pole 6 cm, and the yoke thickness is 4 cm. The septum thickness varies along the magnet, from 4.56 mm at the US end to 2 mm at the DS end of the top pole. The upstream and downstream X centers of the stored beam chamber are separated by 7.887 cm, resulting in a 47.5 mrad rotation of the stored beam chamber in the XZ-plane against the magnet axis. The iron around the stored beam chamber was cut off and made into an open space, as shown in Fig. 2. The material of the stored beam chamber was selected to VP to utilize a higher magnetic permeability to shield the field better than iron.

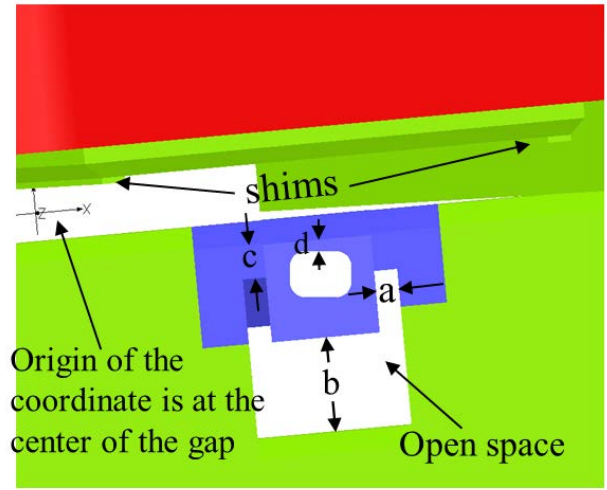

Figure 2: An enlarged view around stored beam chamber at DS

Two shims were applied on both sides of the top pole. The spaces indicated as "a" and "b" in Fig. 2 were optimized to minimize the leakage field inside the stored beam chamber. Dimension "c," in Fig. 2 which shows the wing of the stored beam chamber was tapered up on the right and tapered down on the left from DS to US in order to minimize the skew quadrupole in the stored beam chamber. The septum thickness, d, in Fig. 2 was designed to be 1.4 mm at the end of bottom pole, less than required 2 mm due to the longer bottom pole length and tilt of stored beam chamber.

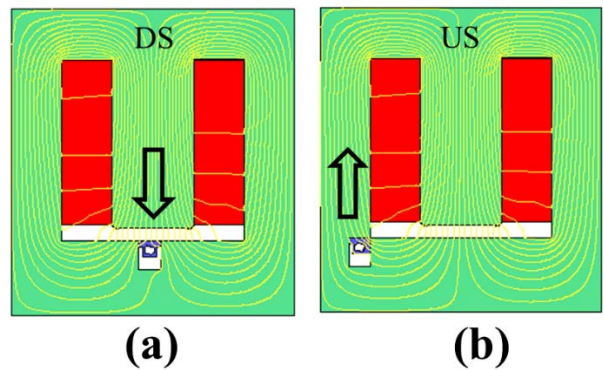

Figure 3: (a) & (b) are the 2D magnetic flux flows at DS and US ends .

The magnetic flux flow at the DS is shown in Fig. 3(a).

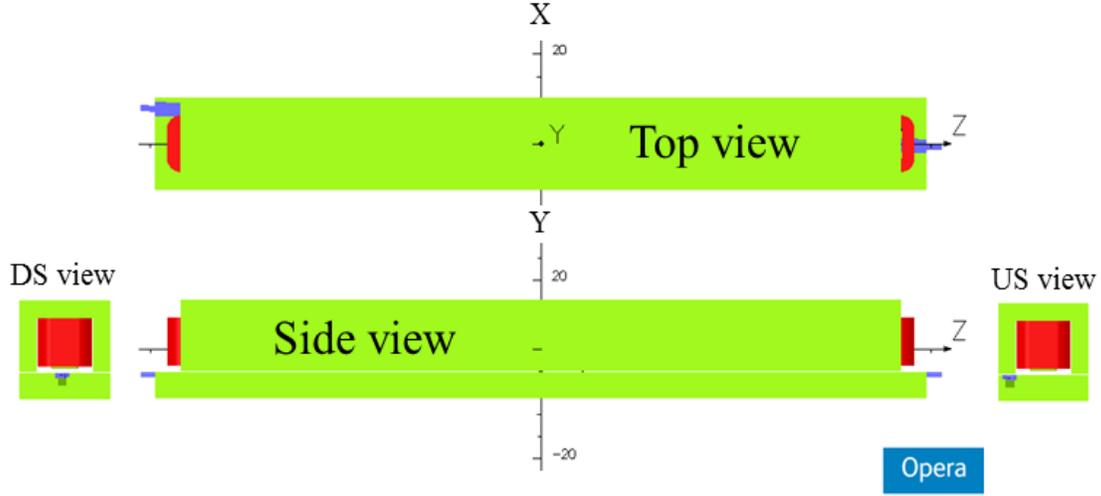

Figure 4: Different views of the septum magnet from the magnet coordinate which is located at the center of the gap in X, Y, and Z. The dimension is in cm.

A strong negative $B_y$ field on top of the stored beam chamber creates a negative $B_y$ leakage field inside the stored beam chamber due to a thin septum of 2 mm at DS. If the US end of the stored beam chamber is placed under the side leg as in Fig. 3 (b) it should create a positive $B_y$ leakage field inside the stored beam chamber. The positive $B_y$ leakage field created in this way should cancel out the leakage field that was created at DS due to a thin septum of 2 mm. Therefore, this concept was applied to the APS-U septum magnet design. The detailed leakage field profiles inside the stored beam chamber with this concept are shown in the result section.

The top pole was designed shorter than the bottom pole as indicated in Fig. 4, in order to reduce the flux density on the bottom pole where the septum is 2 mm.

## RESULTS

To achieve the required field for 89 mrad deflecting angle of the injection beam, a design with 11039 Ampere-turns was chosen. Figure 5 (a) and (b) show the trajectories of the injection and stored beams with a peak field of 1.06 T at the gap center. The trajectories in Fig. 5 (a) refers to the storage ring coordinate system which was set at the US end, and (b) refers to the magnet coordinate which was set at the center of the gap. The outgoing angle, $\alpha_2$, in Fig. 5 (b), was set for the mechanical tilting angle of the stored beam chamber in XZ-plane, which is 47.5 mrad from the magnet axis.

The X position of the injected beam at the DS exit of the septum magnet was set to the center of the stored beam chamber, resulting in the trajectory of the injected beam matching the stored beam trajectory at the DS end as in Fig. 5 (a) and (b).

The total deflecting angle [8] of the injected beam, $\alpha = \alpha_2 - \alpha_1$, was confirmed as -89 mrad from the trajectory in Fig. 5 (b), matching the required angle in the specifications. The $B_x$ and $B_y$ fields along the trajectory of the injection beam were computed and are shown in Fig. 6. The integrated $B_y$ field deflects the injected beam 89 mrad with the ring energy of 6 GeV as required. The integrated $B_x$ field along the injection beam trajectory was decreased to $2.67 \times 10^{-4}$ T - m by optimizing the shims on both sides of the top pole.

Table 2: Integrated multipole fields along the injected beam trajectory (normalized to the dipole)

| | $b_n$ (mm$^{-n}$) = Int_$B_n$ / Int_$B_0$ | $a_n$ (mm$^{-n}$) = Int_$A_n$ / Int_$B_0$ |
|---|---|---|
| 0 | 1 | -1.5e-4 |
| 1 | -2.4e-5 | -7.8e-5 |
| 2 | 5.8e-6 | -5.2e-6 |
| 3 | 2.1e-6 | -1.7e-6 |
| 4 | 2.9e-7 | -4.7e-8 |
| 5 | 3.7e-8 | -6.5e-8 |
| 6 | 9.6e-9 | -1.7e-9 |
| 7 | 1.1e-9 | 2.7e-10 |
| 8 | 3.0e-10 | -4.4e-10 |
| 9 | 9.1e-12 | -1.8e-10 |

The integrated multipole fields, listed in Table 2, are integration of the local multipoles along the beam path. The local multipoles of $B_x$ and $B_y$ fields were calculated by fitting to the equation (1) which is the local field expression in 2D:

$$(B_y + i\,B_x) = \sum_{n=0}^{\infty} (B_n + i\,A_n)\,(x + iy)^n \quad (1)$$

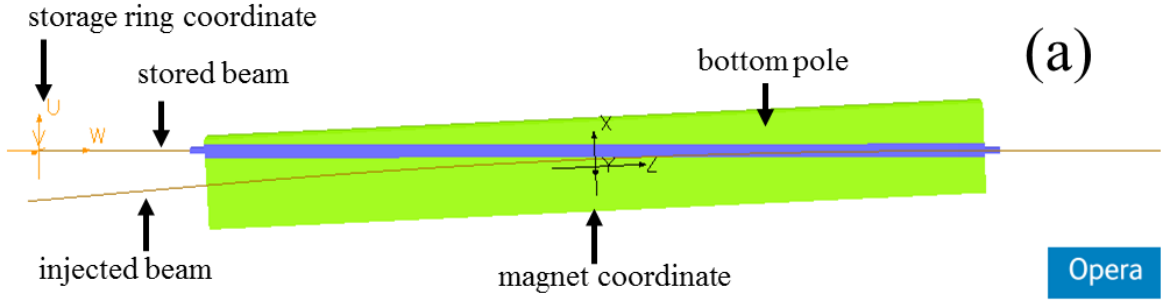

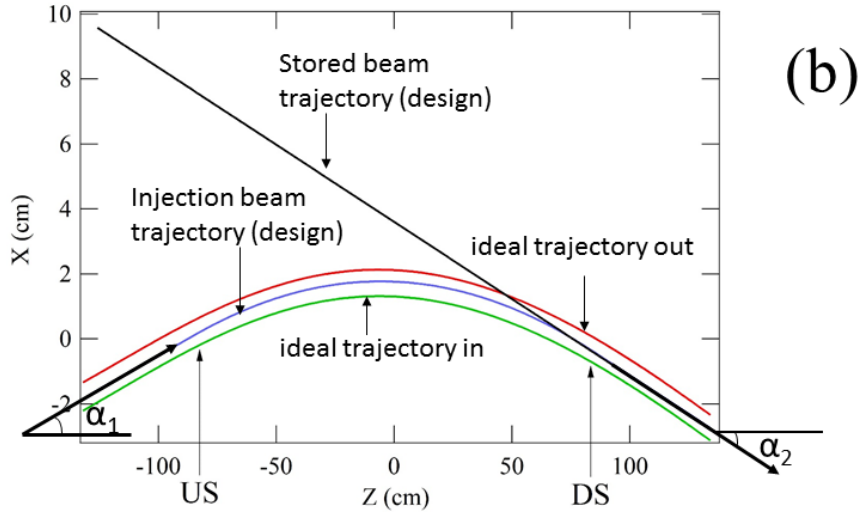

Figure 5: (a) Top views of injected and stored electron beam trajectories of the septum magnet from the storage ring coordinate (all other parts except the bottom pole is hidden); (b) Plots of injected and stored electron beam trajectories in the magnet coordinate system. The ideal trajectory in and out show the allowed range of the injected beam in the XZ-plane using an ideal field.

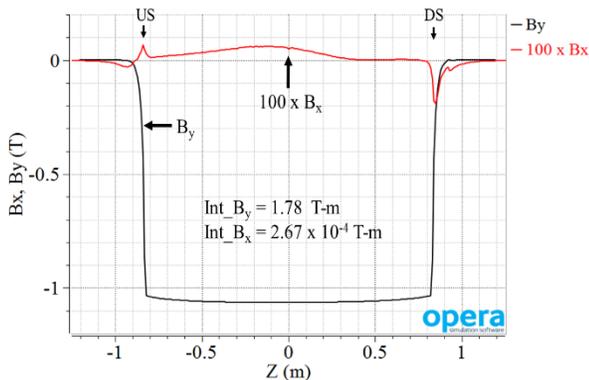

Figure 6: $B_x$ and $B_y$ fields along the injected beam trajectory. The red curve shows the $B_x$ field and the black curve shows the $B_y$ field. The $B_x$ field was multiplied by 100 to show the profile clearly.

Figure 7 shows the local $B_x$ and $B_y$ fields at the center of the stored beam chamber along the length. The peak field is about -10 G at the DS end where the septum is 2 mm. The leakage field at the US was created on purpose.

Figure 8 and 9 show local normal and skew multipole fields in the stored beam chamber from US to DS. As can be seen from Fig. 8, there is a positive normal sextupole field at DS due to the thin septum thickness. However, the negative normal sextupole that was created at US reduced the integrated sextupole over the length.

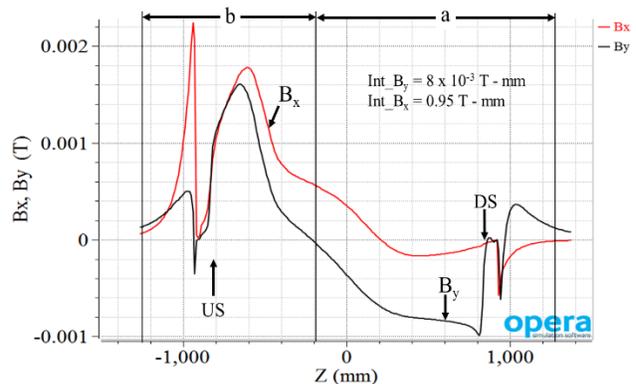

Figure 7: $B_x$ and $B_y$ fields at the center of the stored beam chamber along its length.

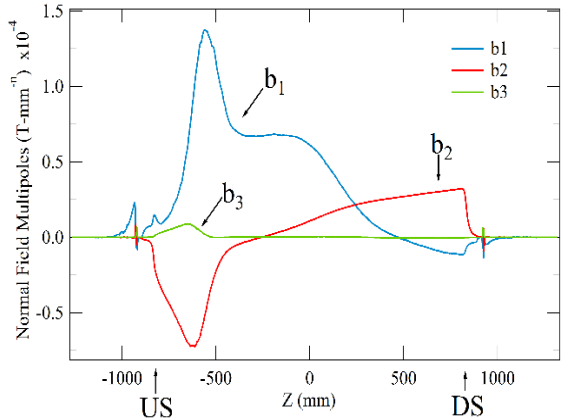

Figure 8: Local normal quadrupole, sextupole, and octupole fields in the stored beam chamber from US to DS.

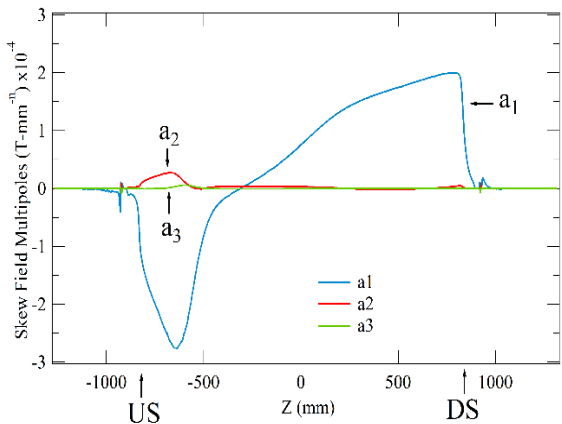

Figure 9: Local skew quadrupole, sextupole, and octupole fields inside the stored beam chamber from US to DS.

Table 3: Integrated multipole fields inside the stored beam chamber

| n | $b_n$ (T-mm$^{-n+1}$) | $a_n$ (T-mm$^{-n+1}$) |
|---|---|---|
| 0 | 8e-3 | 0.948 |
| 1 | 6.65e-2 | 7.61e-2 |
| 2 | 3.53e-3 | 7.07e-3 |
| 3 | 1.69e-3 | 4.1e-4 |
| 4 | 2.46e-4 | -5.6e-4 |
| 5 | -5.1e-5 | 2.7e-5 |

The same thing can be seen from Fig. 9 about the skew quadrupole. The positive skew quadrupole that was created at the DS septum has changed to negative at the US. As a result, the integrated skew quadrupole field was decreased to less than half. The thickness "c" in Fig. 2 was optimized to generate negative skew quadrupole at the US as in Fig. 9, resulting in reduced integrated skew quadrupole. The integrated multipole fields in the stored beam chamber, Table 3, were computed in the area of 2.5 mm in radius at the center of the stored beam chamber along the length.

Figure 10 shows the stored beam displacement in X and Y directions when it passes through the septum magnet from US to DS. The beam enters the septum magnet with zero angle and offset in X and Y at the US (were set), and exits with an offset of 33 µm in X and -75 µm in Y at the DS hard edge (the dashed line in Fig. 10). As can be seen, the beam exits with zero angle (< 1 µrad) in X and -47 µrad in Y. However, the Y angle of the beam at the DS is still a factor of 2 better than the requirement in Table 1.

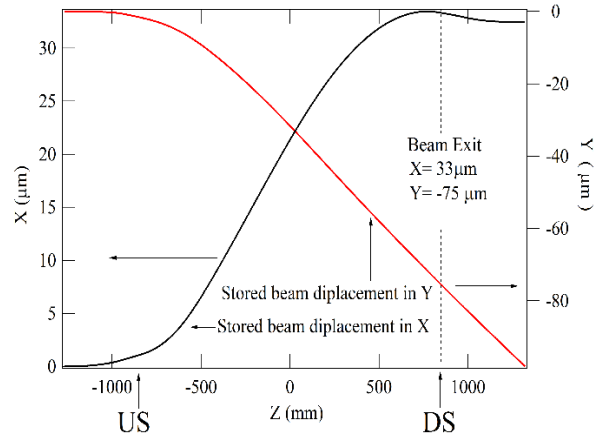

Figure 10: The stored beam displacement in X and Y directions along the beam path.

The field seen by the injected beam was calculated as a function of current and it is shown in Fig. 11. The nominal current is 225.28A to achieve the required deflecting angle, 89 mrad. The magnet efficiency is 88.6% at the nominal current.

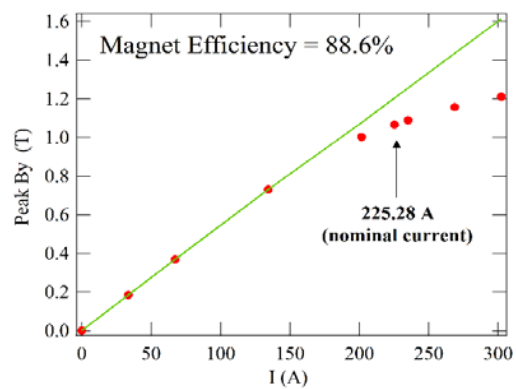

Figure 11: Peak $B_y$ field at the gap center as a function of current. The filled circles are from Opera-3D simulation, and the straight line is a fit to the simulation data at low current, ≤ 150 A.

# DISCUSSION AND CONCLUSION

A septum magnet designed for the APS-U with a 2 mm septum and an injection field of 1.06 T has been presented. The dimensions around the stored beam chamber were optimized to decrease the integrated field leakage in the stored beam chamber. The shims on the pole were optimized to decrease the integrated $B_x$ field along the injected beam trajectory.

A novel concept that cancels out the leakage field was applied to the design. As a result, the integrated negative $B_y$ field in the range "a" in Fig. 7 is cancelled out by the integrated positive $B_y$ field leakage in the range "b". Therefore, the integrated $B_y$ field leakage was only $8 \times 10^{-6}$ T-m in this design. The concept also helped to reduce the integrated normal sextupole and skew quadrupole fields in the stored beam chamber in the way as shown in Fig. 8 & 9.

It is a quite simple but helpful concept to use the returning field at the side leg and create an opposite sign leakage field inside the stored beam chamber.

In parallel to the leakage field cancellation concept, reducing the absolute value of the $B_x$ and $B_y$ leakage fields is critical in order to control the beam displacement when it exits from the septum magnet. The peak $B_x$ and $B_y$ leakage field, especially at DS where the septum is only 2 mm was reduced by: making the top pole shorter than the bottom pole, making an open space around the stored beam chamber, and selecting VP for the material of the stored beam chamber, utilizing its higher magnetic permeability to shield the field better than iron.

These techniques with the tapered VP parts along the length together helped to generate the negative skew quadrupole field at US which resulted in compensation of the positive integrated skew quadrupole in the stored beam chamber at the DS end.

The calculated second field integrals (beam offset) showed that there is a 33 µm and -75 µm displacement of the stored beam in horizontal and vertical, respectively, when it exits from the septum magnet.

# REFERENCES

[1] M. Borland, V. Sajaev, and Y.P. Sun, "Lower Emittance Lattice for the Advanced Photon Source Upgrade Using Reverse Bending Magnets," presented at NAPAC'16, Chicago, USA, October 2016, paper WEPOB01.

[2] G. Decker, "Design Study of an MBA Lattice for the Advanced Photon Source," Synchrotron Rad. News, vol. 27, No. 6, 13-17 (2014).

[3] R. Abela et al., EPAC92, 486-488 (1992).

[4] L. Emery et al., PAC03, 256-258 (2003).

[5] A. Xiao, IPAC15, 1816-1818 (2015).

[6] S. Sheynin, F. Lopez and S.V. Milton, "The APS Direct-Drive Pulsed Septum Magnets" IEEE1996, 1355-1357 (1996).

[7] J. Rank, G. Miglionico, D. Raparia, N. Tsoupas, J. Tuozzolo, Y.Y. Lee, "The Extraction Lambertson Septum Magnet of the SNS" PAC2005, 3847 -3849 (2005).

[8] H Wiedemann, "Particle Accelerator Physics II" Springer, p135, Berlin (1995)